\documentclass{PoS}

\title{Magnetic Reconnection on Jet-Accretion disk Systems}

\ShortTitle{Magnetic Reconnection on Jet-Accretion disk Systems}

\author{\speaker{Elisabete M. de Gouveia Dal Pino}\\
       Instituto de Astronomia, Geof\'isica e Ciencias Atmosfericas - Universidade de S\~{a}o Paulo (IAG/USP), S\~{a}o Paulo, Brazil\\
        E-mail: \email{dalpino@iag.usp.br}}

\author{Maria Victoria del Valle\\
        Instituto Argentino de Radioastronomıa (IAR), Argentina\\}
\author{Luis Kadowaki\\
        IAG - Universidade de S\~{a}o Paulo, Brazil\\}

\author{Behoruz Khiali\\
        CERN, Switzerland\\}

\author{Grzergorz Kowal\\
        EACH - Universidade de S\~{a}o Paulo, and UniSul, Brazil\\}

\author{Yosuke Mizuno\\
       Institute for Theoretical Physics, Goethe University, Germany\\}

\author{Chandra B. Singh\\
       IAG - Universidade de S\~{a}o Paulo, Brazil\\}

\abstract{Fast Magnetic Reconnection is currently regarded as an important process also beyond the solar system, specially in magnetically dominated regions of galactic and
extragalactic sources like  the surrounds of black holes and relativistic jets. In this lecture we discuss briefly the theory of fast magnetic reconnection, specially when driven by turbulence which is very frequent in Astrophysical flows, and its implications for relativistic particle acceleration. Then we discuss  these processes  in the context of the sources above, showing recent analytical and multidimensional numerical MHD studies that indicate  that fast reconnection  can be a powerful process to accelerate particles to relativistic velocities, produce the associated high energy non-thermal emission, and account for  efficient conversion of magnetic into kinetic energy in these flows.}

\FullConference{Frontier Research in Astrophysics II\\
		23-28 May 2016\\
		Mondello (Palermo), Italy}

\begin{document}

\section{Introduction}
Black Holes (BH) and the associated accretion and relativistic jet phenomena are  ubiquitious in Astrophysics. There are evidences for the presence of 
stellar-mass BHs in galactic binary systems,  usually named GBHs (or BHBs, or microquasars). Also, supermassive BHs (with   $\sim 10^6$ to $10^{10}$  solar masses) are believed to occur in the nuclei  of all classes of active galaxies (or AGNs, from the high luminous  blazars to the less luminous   radio or  seyfert galaxies). Besides, it is also argued that BHs may be the ultimately engines of gamma-ray-burts (GRBs).

Multi-wavelength observations  indicate that the collimated relativistic jets produced near the central BH can be accelerated to large Lorentz factors,  up to several orders of magnitude in length scales. Their formation is still a matter of debate, but the usually  most accepted models rely on magnetic processes, like  magneto-centrifugal acceleration by helical magnetic fields arising from the accretion disk \cite{blandford_payne82}. Or otherwise  they can be powered by the BH spin transferred to the surrounding magnetic flow  \cite{Blandford_Znajek1977}. 
Thus in any case the  prediction is  that the jets should be  born as magnetically  or  Poynting flux dominated flows. The fact that at  observable distances from the source (starting  around $1000 R_S$ or less, where $R_S$ is the Schwartzschield radius) these jets become kinetically dominated, indicate that they must somehow suffer an efficient  conversion (or dissipation) of magnetic into kinetic energy. We  argue below that magnetic reconnection may be a powerful mechanism operating in these jets to allow for this conversion. 

Another problem that is currently  challenging the researchers regards  the origin of the very high energy (VHE) emission of these sources. For instance,  until recently only AGNs with highly beamed jets pointing to the line of sight, namely  blazars,   were detected by gamma-ray telescopes.  More than a  chance coincidence, these detections  are consistent with the conventional scenario that attributes the VHE emission of these sources to particle acceleration along the jet being strongly Doppler boosted and producing  apparently very high fluxes. Recently, however, a few non-blazar sources which belong to the branch of low luminosity AGNs (or simply LLAGNs)  for having   bolometric luminosities of only a few times the Eddington luminosity, 
%$L_{Edd}$ 
%\cite{nagar_etal_05}
%\citep[see][]{zhang_yan_11, spitzer_62}.
have been also detected at TeV energies by  ground based gamma-ray observatories (e.g. \cite{sol13} and references therein). Among these sources, the radio galaxies M$87$, Centaurus A (or Cen A), Persus A (or Per A), and IC 310 are the most striking examples.
The angular resolution and  sensitivity of these detectors are still  so  poor that it is hard to   establish  if it comes from the jet or from the core.
 %(e.g., \citealt{kachel10}). 
These VHE detections were surprising because, besides being highly underluminous, the viewing angle of the jets of these sources is of several degrees, therefore allowing for only moderate Doppler boosting. These  characteristics make it hard to explain the VHE of these sources adopting  the same standard scenario of blazars. Furthermore, 
observations of  short time scale variability in the gamma-ray emission  of  IC 310, M87 and Per A  
 (e.g., \cite{ahar06,abdo09,ackermann12, aleksic14} and references therein) indicate that it  is produced in  a very compact region that could be perhaps the core. 
 In the case of Cen A, though  there is no evidence of significant variability at VHE, it has been also argued that   the TeV data of this source would be more compatible with a point source near the core  \cite{kachel10}. 
%If the gamma-ray photons were due, for instance,  to proton-proton  interactions along the jet then on leaving the source they would interact with the extragalactic background light (EBL) resulting in a flatter spectrum in the TeV range than the currently measured  with \textit{HESS} \cite{kachel09b}.

Though a number of works have attempted to explain these observations as produced in the  jets of these sources (e.g. \cite{tavecchio08, lenain08, gian10}),  the evidences above led to the search for alternative particle acceleration scenarios involving the production of the VHE in the surrounds of the  BH, in the core region (see e.g.  \cite{kadowaki15, khiali16a} and  references therein).   If such environments are magnetically dominated as generally supposed, then fast magnetic reconnection may be unavoidable \cite{gl05, degouveia15}.

Researchers have been facing  similar difficulties also  with the interpretation of the  VHE emission  of GBHs and their jets, particularly with the sources Cygnus X1 and X3 for which  recent detections indicate upper limits in the TeV range too  \cite{kadowaki15, singh15,khiali15}. 

Even in the cases where the emission is most probably produced along the relativistic jets, as in blazars,  the conventional mechanism to accelerate the particles based on diffusive shock acceleration is facing severe constraints imposed by current VHE observations with very  high variability, of the order of  minutes in the TeV range. This implies extremely compact acceleration/ emission regions ($< R_S/c$) and besides,  it requires bulk Lorentz factors larger than the typical  values expected for such sources (around $\gamma \simeq 5-50$)  in order to avoid electron-positron pair creation. To circumvent these problems,  \cite{giannios09} proposed a reconnection model involving misaligned $mini-jets$ inside the jet. Fast reconnection has been also invoked in \cite{zhangyan11} to explain the transition from magnetically to kinetically dominated flow and the prompt gamma-ray emission in GRBs.

In this lecture, we discuss the role of fast magnetic reconnection in  accretion disk/jet systems around BHs  and show that this can be a powerful process to accelerate relativistic particles, produce the associated non-thermal emission and  particularly  the VHE one, and account for an efficient conversion of magnetic into kinetic energy.

\section{Fast Magnetic Reconnection and Particle Acceleration}

Fast magnetic reconnection occurs when two magnetic fluxes of opposite polarity encounter each other and partially annihilate at an efficient rate $V_R$ close to the local Alfv\'en speed ($V_{A}$). Besides successful laboratory reconnection experiments \cite{yamada_etal_10} and direct observations in the earth magnetotail and solar flares,   extensive  numerical work has been also carried out to understand the nature of this process   both in collisionless (e.g., \cite{birn01}) and collisional plasmas (e.g., \cite{kowal09,loureiro07}). Different mechanisms such as plasma instabilities, anomalous resistivity,
%(e.g., \cite{biskamp97,zenitani09}),  
and turbulence, can lead to fast reconnection. The latter process has been found to be   very efficient and probably the main driving mechanism of fast reconnection in collisional MHD flows \cite{LV99}.  
Even weak embedded turbulence causes the wandering of the magnetic field lines which allows for many independent patches to reconnect simultaneously making the  reconnection rate  large and independent on the local microscopic magnetic resistivity, $V_R \sim v_A (l_{inj}/L)^{1/2} (v_{turb}/v_A)^2$, where  $l_{inj}$ and  $v_{turb}$ are the injection scale and velocity of the turbulence, respectively \cite{LV99}. 
One should thus expect that in magnetically dominated media the release of energy will
result in  outflow  motions that in high Reynolds number media
 drive turbulence that will in turn speed up the reconnection further.
%accelerating the energy release. 

The break of the magnetic field topology by fast reconnection involves the  release of  magnetic energy
explosively which explains, for instance, the bursty emission in  solar
flares. Relativistic particles are always observed in connection with these
flares suggesting that magnetic reconnection can lead to direct particle
acceleration (see for a review \cite{degouveia15}).

In analogy to diffusive shock acceleration (DSA), in which particles confined
between the upstream  and downstream  flows undergo a first-order Fermi
acceleration, \cite{gl05} (hereafter GL05) proposed  a similar
process 
 where trapped particles bounce
back and forth between the converging magnetic fluxes   of a 
large scale reconnection discontinuity (or current sheet). The particles  gyrorotate around a reconnected
magnetic field (see Figure 2b in \cite{kowal11}, gaining energy due to
collisions with magnetic irregularities at a rate $\Delta E/E \propto V_{R}/c$  {implying} a
first-order Fermi process with an exponential energy growth after several round
trips (GL05).  This process has been extensivelly tested numerically mainly through two-dimensional  particle-in-cell (PIC) simulations of collisionless  plasmas \cite{drake06,zenitani01,zenitani07, zenitani08,lyubarsky08,drake10,clausen-brown2012, cerutti14, li15}, and more recently also through three-dimensional (3D) PIC simulations \cite{sironi14,guo15, guo16}. However, these simulations can probe acceleration only at the kinetic scales of the plasma, of a few hundreds of the inertial length   ($\sim 100 c/\omega_p$, where $\omega_p$ is the plasma frequency). To assess the first-order Fermi process in the large scales of the collisional MHD flows commonly observed in astrophyisical systems, \cite{kowal11, kowal12, delvalle16} have also successfully tested it in 2D and 3D MHD simulations injecting thousands of test particles in the reconnection domain.
 
Figure 1 depicts a typical trajectory of a test particle in a large scale 3D MHD magnetic discontinuity where turbulence was embedded to make reconnection fast \cite{kowal12, delvalle16}. 
These simulations  indicate  an efficient particle acceleration rate $t_{acc}^{-1} \propto E^{-\alpha}$, with $0.2 < \alpha < 0.6$ for a vast range of values of  $c / V_{\rm A} \sim 20 - 1000$, and an accelerated particle  power-law spectrum  which in the initial times of the simulation can be fitted by $N(E)\propto E^{-1,-2}$, where $E$ is the particle kinetic energy \cite{delvalle16}. 

\begin{figure}[!h]
\begin{center}
\includegraphics[angle=0,scale=0.32]{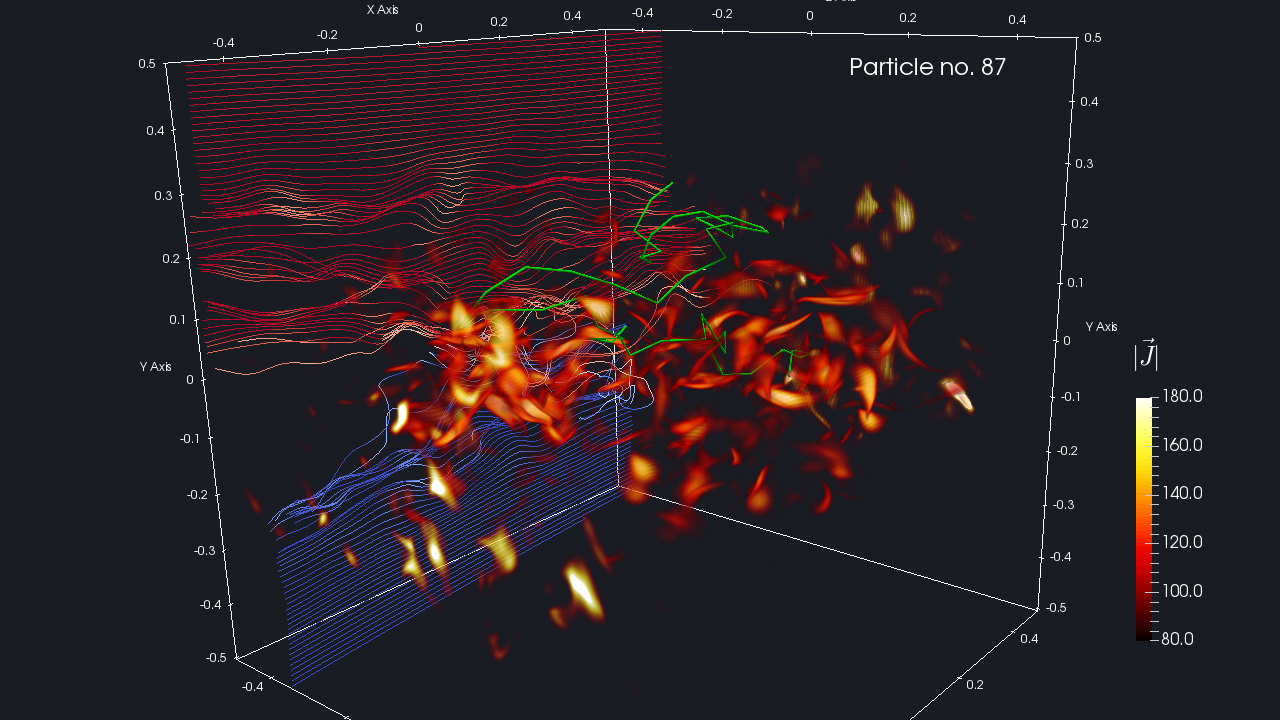}\\
\end{center}
\caption{Map illustrating the trajectory (in green) of an injected test particle bouncing up and down in the  3D magnetic reconnection domain simulated with  turbulence embedded in it  to make reconnection fast. The volume depicts the current sheet region  with patches of very high current density ($J$) where the particle is scattered by the magnetic fluctuations while accelerated. The blue and red lines represent the two magnetic line fluxes of opposite polarity (extrated from \cite{delvalle16}).
}
\label{evo-maps}
%\end{centering}
\end{figure}

Currently, fast magnetic reconnection is regarded as an important
mechanism to accelerate particles also beyond the solar system,  in magnetically dominated regions of galactic and
extragalactic sources like pulsars (e.g. \cite{clausen-brown2012,cerutti14}), jets and the surrounds of BHs   (e.g. \cite{gl05,degouveia10a,degouveia10b, giannios10,delvalle11,
kadowaki15,khiali15,khiali16a, zenitani07, sironi14, guo15,singh16}), so that the information above obtained from the numerical simulations is  crucial for modelling the non-thermal emission  produced out of accelerated relativistic particles in such sources.

In the following sections, we will discuss applications of this acceleration mechanism driven by magnetic reconnection  in the framework of BH systems and relativistic jets.

%FIGURE 1

\section{Magnetic Reconnection in the surrounds of BHs}

In earlier work (GL05) \cite{gl05}, we discussed a model in which particles can be accelerated through  the first-order Fermi process described above, in the surrounds of a BH, in the inner coronal region, by the  power extracted from fast magnetic reconnection events occurring between the field lines of the magnetosphere of the BH  and the  field lines arising from the accretion disk (see Figure 2),  based on  similar phenomena occurring  in the solar corona. This model was first explored in the framework of microquasars and then  extended to AGNs  \cite{degouveia10a,degouveia10b}.   
These works revealed  that fast reconnection could be efficient enough to produce the core radio outbursts in microquasars and AGNs. For a detailed description of the model see \cite{kadowaki15}.

We should remark that recent global non-relativistic (e.g.  \cite{romanova11, kadowaki11, zanniferreira13}) and general relativistic \cite{dexter14} MHD  numerical simulations of accretion disk systems have evidenced magnetic reconnection events in the coronal region around the central source.

% FIGURE 2
\begin{figure}[!h]
\begin{center}
\includegraphics[angle=0,scale=0.28]{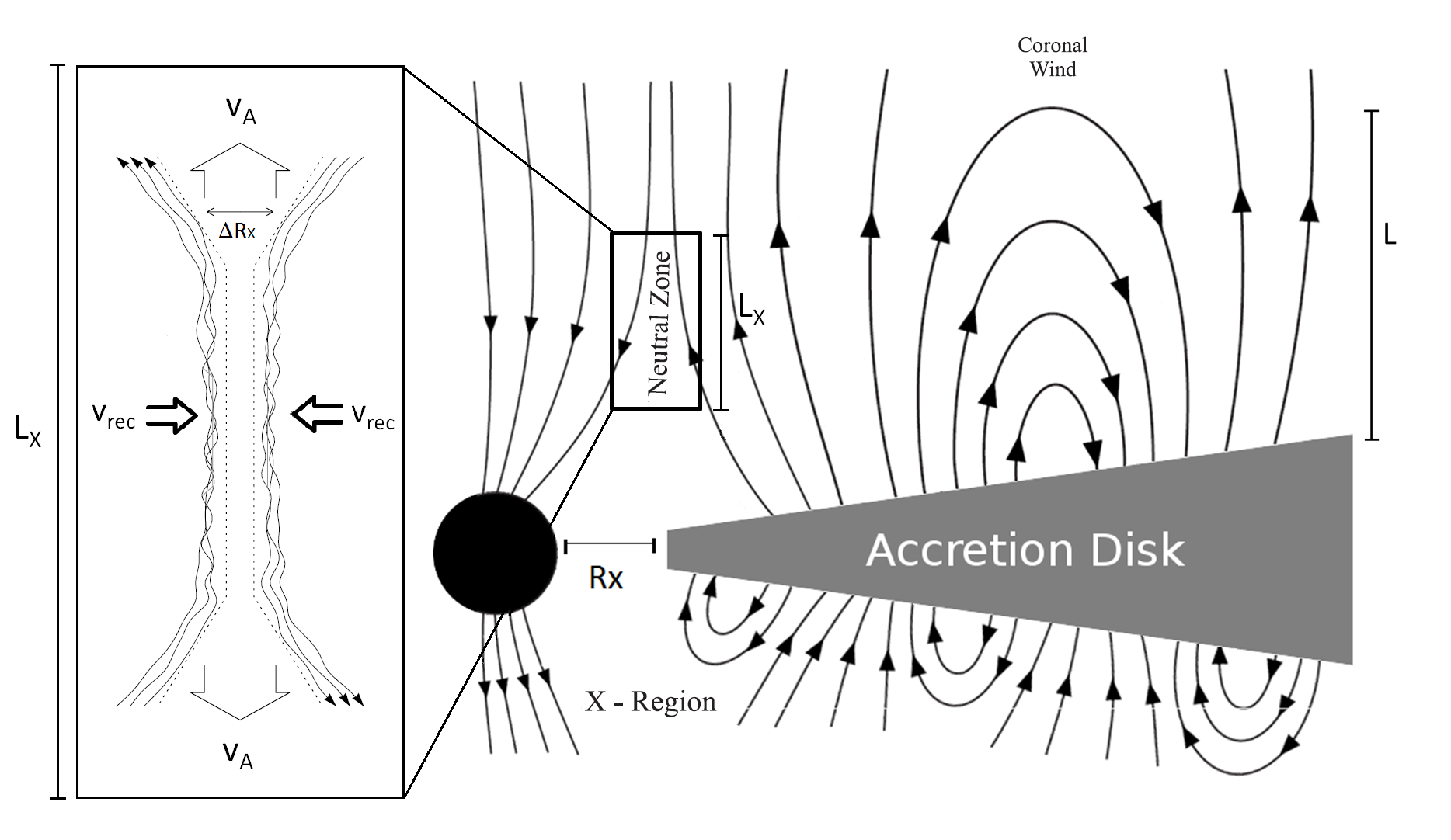}\\
\end{center}
\caption{
Idealized scheme of magnetic reconnection between the lines of the BH magnetosphere and those emerging from the accretion disk into the corona. Reconnection is made fast by the presence of embedded turbulence in the reconnection zone (as indicated in the detail) and allows for extraction of  large amounts of reconnection power (eq. 1). Particle acceleration may also naturally occur in the magnetic reconnection zone by a first-order Fermi process (adapted from \cite{kadowaki15, khiali16a}). This representation is suitable for a geometrically thin disk, but we obtain similar results when considering a magnetized ADAF-type accretion disk  \cite{singh15}. }
% \label{figaccretionBH}
\end{figure}

More recently,  the model above has been revisited  to explore the effects of   different mechanisms of fast magnetic reconnection and accretion  and to investigate the origin of the  gamma-ray emission of  a large sample of BH sources  including AGNs, GBHs  and even GRBs \cite{kadowaki15, singh15}. 

The  magnetic reconnection power released   by  fast reconnection in the magnetic discontinuity region as schemed in Figure 2 is given by \cite{kadowaki15}:

\begin{equation} \label{2}
W \simeq 1.66\times 10^{35} \psi^{-0.5} r_X^{-0.62} l^{-0.25} l_X q^{-2}\dot{m}^{0.75}m\ ~ {\rm erg~s^{-1}},
\end{equation}
where $m$ is the mass of the BH in units of solar mass, $\dot{m}$ is the accretion rate in units of the Eddington accretion, $r_X$ is the disk inner radius normalized by $R_S$,  $l=L/{R_S}$ is the height of the corona in units of $R_S$; $l_X={L_X}/{R_S}$, $L_X \leq L$ is the extension of the magnetic reconnection zone (as shown in Figure 2), $q=[1-(3/r_X)^{0.5}]^{0.25}$, and $\psi=[1+({v_{A0} \over c})^2]^{-1/2}$, with $v_{A0} = B/(4\pi \rho)^{1/2}$,  $B$ being the local magnetic field,  $\rho \simeq n_c m_p$  the fluid density  in the corona, $n_c$ the coronal  number density, and  $m_p$ the proton mass. This equation has been derived considering the natural presence of turbulence in the current sheet as the driving source of  fast reconnection, and taking  a geometrically  thin accretion disk model \cite{kadowaki15}.

Figure 3 depicts this power as a function of the BH mass (gray region), considering  a fiducial parametric space ($ 0.0005 \leq \dot{m} \leq 1$, $r_X=6$, $1 \leq l  \leq 18$, $l_X \leq l$, and $\psi \simeq 1$) \cite{kadowaki15}.   The results of this study  confirmed a  trend predicted earlier  in \cite{degouveia10a, degouveia10b}  that  there is a correlation between the calculated fast magnetic reconnection power and the BH mass spanning $\sim$ 10 orders of magnitude. This can explain
not only the observed radio, but also the gamma-ray emission from GBHs
and low luminous AGNs (LLAGNs). This match is found for the emission of more than 230 sources (indicated in the diagram with red and green symbols) which include those of the so called
fundamental plane of black hole activity \cite{merloni03}. This figure also reveals  that the observed emission from blazars
 and GRBs
does not follow the same trend as that of the LLAGNs and GBHs,
indicating  that the observed radio and gamma-ray emission in these cases is
not produced in the core of these sources. This result is actually not a surprise,  because the jet in these systems points to the line of sight and
thus screens the nuclear region, so that in these sources the emission is
expected to be produced by another population of accelerated particles, along
the jet.

%FIGURE 3

\begin{figure}[!h]
\begin{center}
\includegraphics[angle=0,scale=0.85]{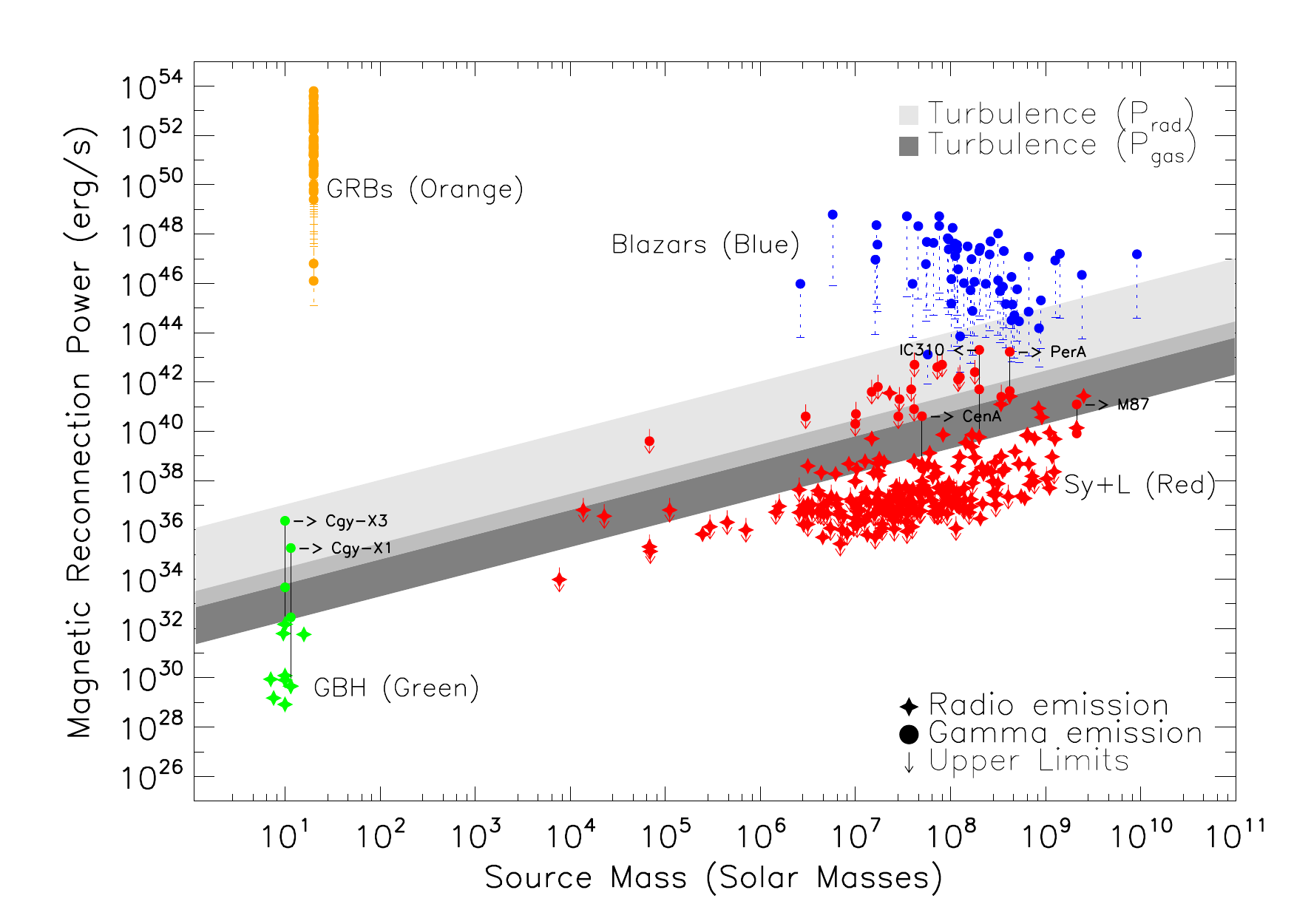}
\end{center}
\caption{
Calculated turbulent driven magnetic reconnection power versus BH source mass (gray
region) compared to the observed  emission of low luminous AGNs (LLAGNs: LINERS
radio-galaxies, and Seyferts), galactic black hole binaries (GBHs), high luminous AGNs (blazars)
and gamma ray burst (GRBs). The core radio emission of the GBHs and LLAGNs is
represented by red and green diamonds, the gamma-ray emission of these two
classes is represented by red and green circles, respectively.  In the few cases
for which there is observed gamma-ray luminosity it is plotted the maximum and
minimum values linking both circles with a vertical black line that extends down
to the radio emission of each  of these sources. The inverted arrows associated
to some sources indicate upper limits in gamma-ray emission.  For blazars and
GRBs, only the gamma-ray emission is depicted, represented in blue and orange
circles, respectively. The vertical dashed lines correct the observed emission
by  Doppler boosting effects. The calculated reconnection power  clearly matches
the observed radio and gamma-ray emissions from LLAGNs and GBHs, but not that
from blazars and GRBs. This result confirms early evidences that the emission
in blazars and GRBs is produced along the jet and not in the core of the sources, but on the other hand do indicate that the gamma-ray emission of LLAGNs and GBHs can be instead produced in the core (extracted from \cite{kadowaki15}). }
%\label{figAGN-BHB}
\end{figure}

In another concomitant  work \cite{singh15}, we  explored the same
mechanism, but instead of considering a standard thin, optically thick accretion
disk as in the  works above, we adopted a  magnetically-dominated advective
accretion flow (M-ADAF, \cite{narayanYi95, meier12})  around the BH, which
may be more suitable for sub-Eddington sources. The results   obtained are very similar
to those of \cite{kadowaki15} depicted in  Figure 3, suggesting
 that the details of the accretion physics are not affecting much the
turbulent magnetic reconnection process which actually occurs in the $corona$
around the BH and the disk.

The correlations found  in Figure 3 \cite{kadowaki15, singh15} have encouraged further testing of  magnetic reconnection  around BH sources.  Employing  the
reconnection induced particle acceleration model described above  and
considering the relevant non-thermal  loss processes  of the accelerated
particles  (namely, Synchrotron, inverse Compton, proton-proton and
proton-photon processes), we   computed the spectral energy distribution (SED) of
several GBHs \cite{khiali15} and LLAGNs \cite{khiali16a} and found that these match very well with the
observations. As an example,    Figure 4 depicts the calculated SED of
the radio-galaxy M87 compared with the observed fluxes  from radio to gamma-rays. These results  strengthen the
conclusions above in favour of a core emission origin for the very high energy
 of these sources. The model also naturally  explains  the observed high
 variability of the emission which is produced in a very compact region. 
 We should also remark that we have  calculated the absorption  of the calculated gamma-ray emission via  pair creation due to interactions with  photons of the broad line region (BLR)    of this source and found it to be negligible since  its axis angle to the line of sight is around 30 degrees (rather than being edge-on).

The same acceleration model has been also recently applied
to  explain the high energy extragalactic neutrinos observed  by the IceCube at TeV energies. Based on the results of  \cite{khiali16a},  we have shown that  protons accelerated in the core  region of  radio galaxies distributed between redshifts 0 and  $\sim 5.2$,  can  produce the observed flux of  neutrinos via the decay of charged pions produced by the photomeson process suffered by accelerated protons in these sources \cite{khiali16b}.

\begin{figure}
\begin{center}
\includegraphics[angle=0,scale=0.85]{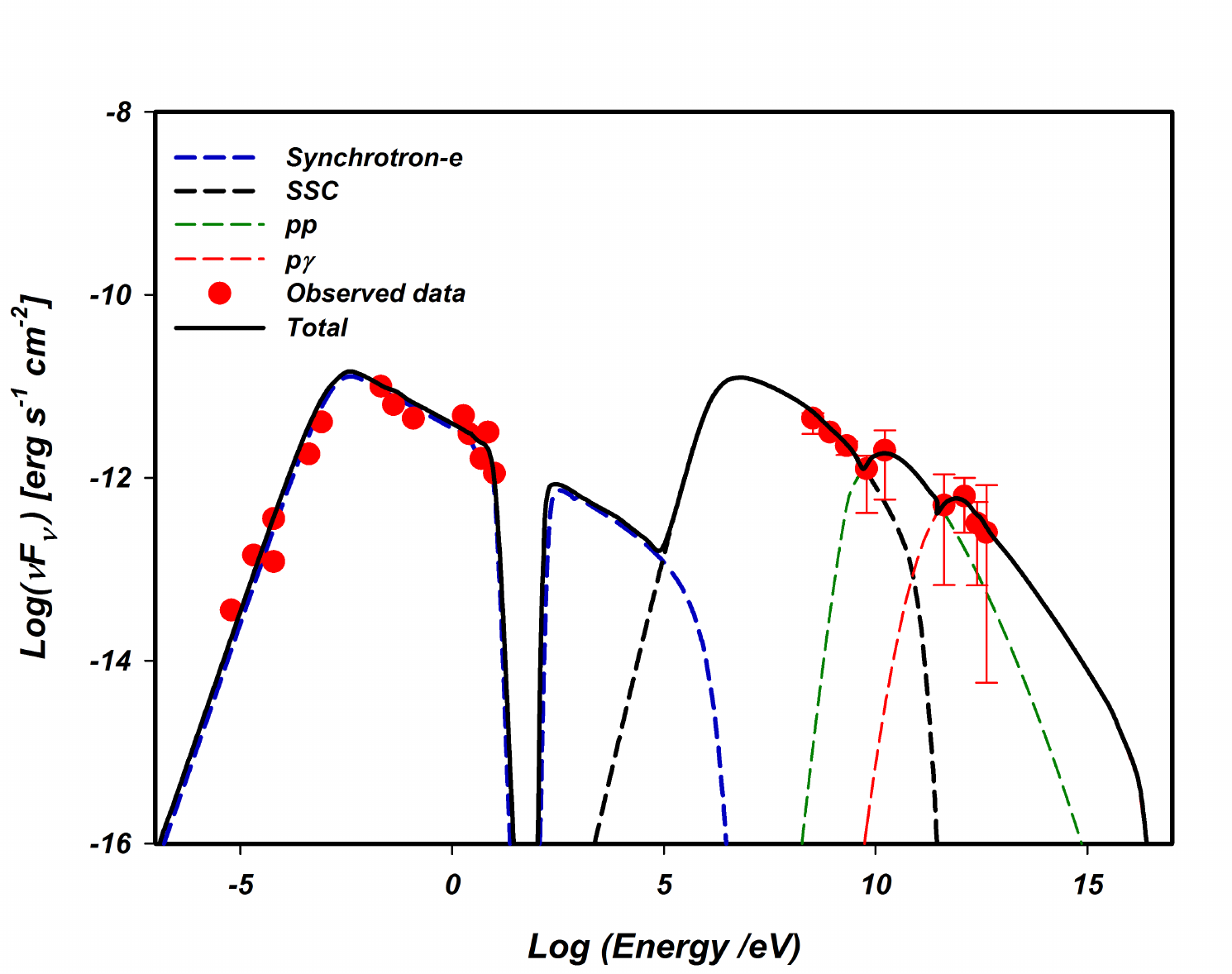}
\end{center}
\caption{Calculated spectral energy distribution (SED) for the AGN M87
employing  the turbulent magnetic reconnection acceleration model in the core
region. The core radio data are obtained from MOJAVE VLBA at 15 GHz,  \cite{biretta91}  at 1.5, 5 and 15 GHz,  IRAM at 89 GHz,  SMA at 230 GHz,   Spitzer at 21 and 7.2 GHz, and  \cite{perlman01}  at 3.2 GHz; optical-UV emission from HST,  \rm{MeV/GeV} $\gamma$-ray   from \textit{Fermi-LAT}; and the  low-state TeV spectrum  from \textit{HESS} 
 (extracted from  \cite{khiali16a}; see also references therein).
 }
%\label{SED-M87}
\end{figure}

\section{Magnetic Reconnection along Relativistic Jets}

As remarked in Section 1, fast  magnetic reconnection has been  also lately invoked  to explain the dynamical processes involved in effective conversion of magnetic  into kinetic energy and on particle acceleration along relativistic jets, specially of AGNs and GRBs  (for  recent reviews see  \cite{mizuno16, lazarian16} and references therein).

\begin{figure}[!h]
\begin{center}
\includegraphics[angle=0,scale=1.05]{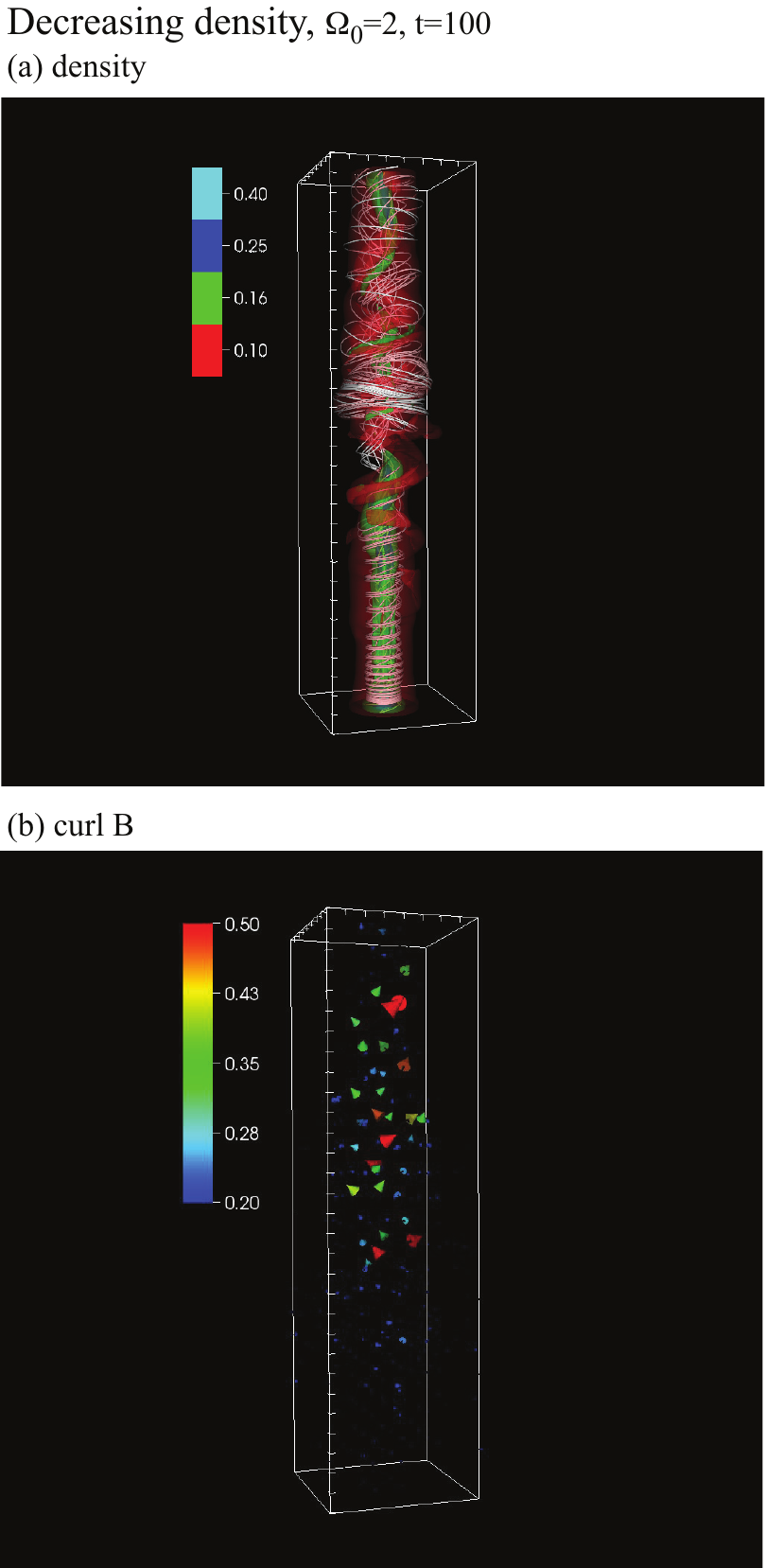}
\includegraphics[angle=0,scale=0.15]{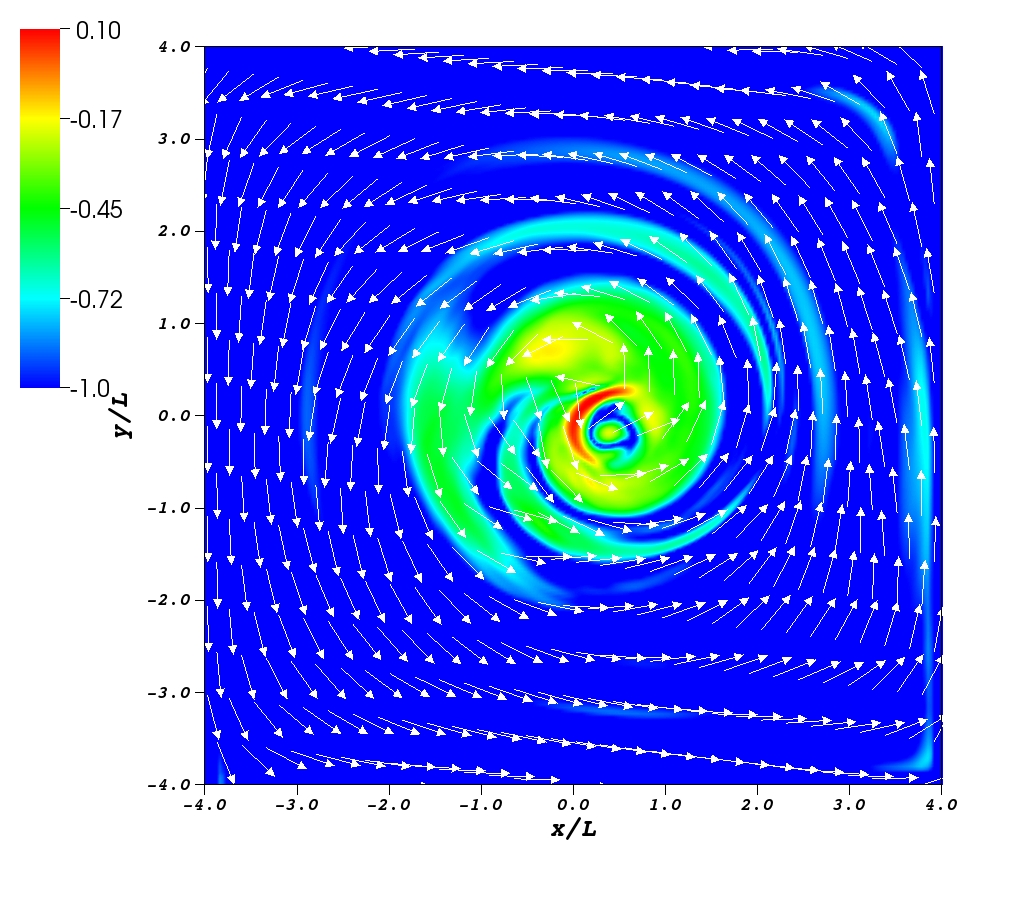}
\end{center}
\caption{Current driven kink instability and fast magnetic reconnection in a relativistic MHD jet. The upper diagram depicts 3D density isosurfaces,  and the bottom left diagram
the locations of maximum current density  which trace  magnetic reconnection sites. The bottom right diagram depicts  a  transverse cut  at z = 13  showing  the
logarithm of the current density  with (white) vectors of the magnetic field superposed to it. We note clearly a large reconnection sheet near the axis (in red)  which coincides with a region of minimum magnetization  ($\sigma$) in the flow due to energy dissipation (see details in \cite{singh16}).
}
% \label{f10_curlB_dec}
\end{figure}

At the inner scales, around  $1000 R_S$ from the central source or less, these jets are expected to be Poynting flux dominated flows and as such, are probably loci of strong helical fields  driven by rotation at the jet launching region. Jets with strong toroidal fields are in turn susceptible to current-driven kink (CDK) instability (e.g. \cite{mizuno12}).  This  excites large-scale helical motions that can  strongly distort or even disrupt the beam, and induce reconnection.

\begin{figure}[!h]
\begin{center}
\includegraphics[angle=0,scale=1.0]{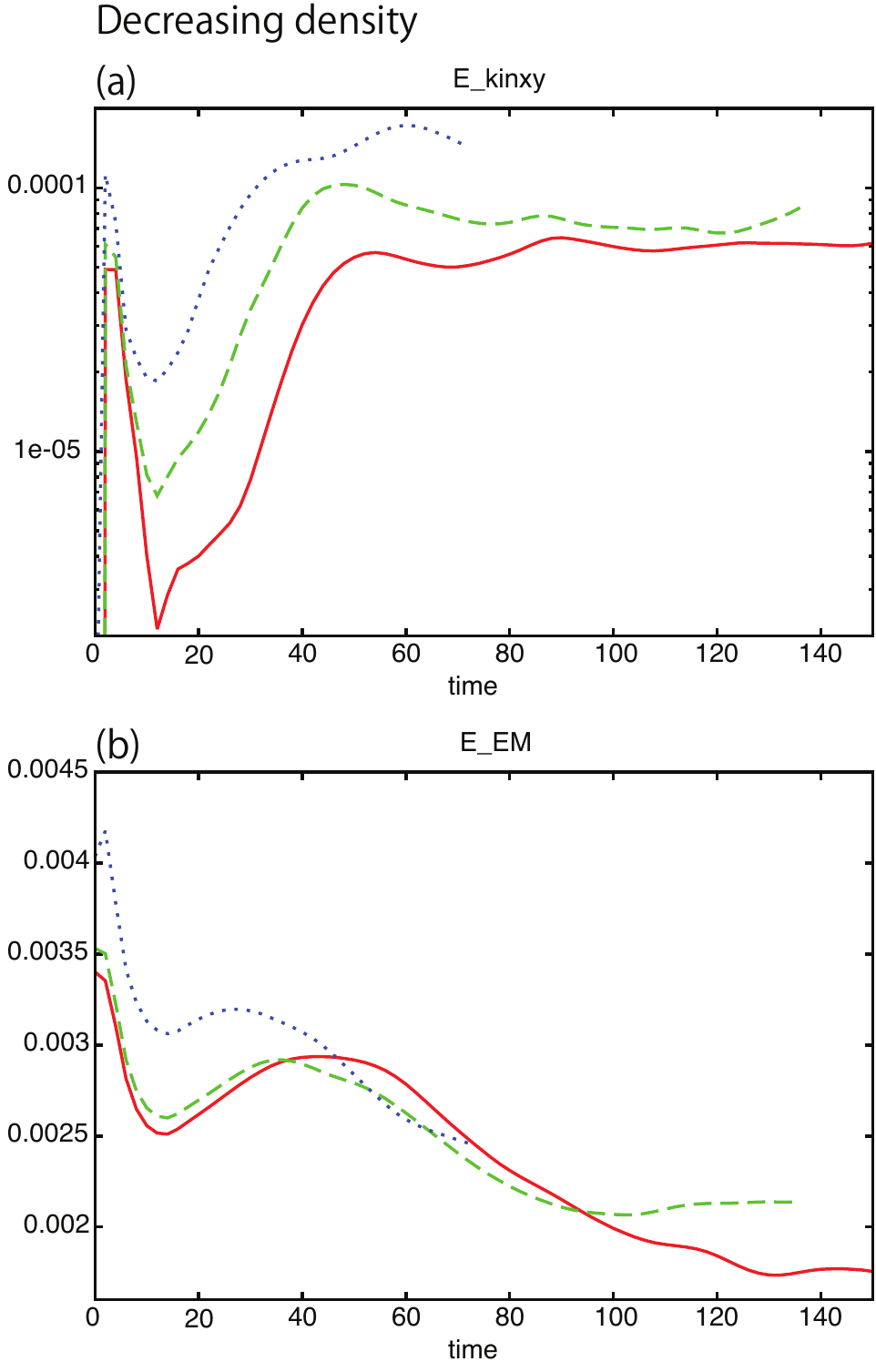}
\end{center}
\caption{Volume-averaged time evolution of the kinetic (top) and magnetic (bottom) energies for the
jet model presented in Figure 5 with angular velocity $\Omega = 2$ (dashed green line) compared to other two models with  $\Omega = 1$ (solid red line), and $\Omega = 4$ (blue dotted line). The depicted  kinetic energy  is  transverse to the z-axis, and the magnetic energy  is the total integrated relativistic electromagnetic energy (extracted from  \cite{singh16}).
}
% \label{f10_curlB_dec}
\end{figure}

To probe this process, we have  performed recently  3D relativistic MHD simulations of rotating Poyinting flux dominated tower jets with initial helical fields.  Considering models with a ratio between the magnetic and the rest mass energy of the flow $\sigma\simeq 1$,  and different  density ratios between the jet and the environment, we induced  precession perturbations that quickly developed  CDK modes. Figure 5 shows an example of a jet with a density larger than that of the environment.  The results reveal the propagation of a helically kinked structure along the jet that causes substantial dissipation of magnetic  into kinetic energy \cite{singh16}. Connected to this, we identify regions of maximum current density that trace filamentary current sheets, where fast magnetic reconnection driven by the CDK instability takes place  with rates $\sim 0.05 V_A$. This value is consistent with  recent studies of fast turbulent reconnection  in the  relativistic regime \cite{takamoto15} which in turn predict similar reconnection rates as those in non-relativisitc flows \cite{LV99}. Figure 6 depicts the time evolution of the magnetic and kinetic energy of the jet flow of Figure 5 highlighting the dissipation of the first in favour of  the increase of the second. We also find that the reconnection regions are directly correlated with zones of decreasing sigma  in the flow. 

The implications of these findings for Poynting-flux dominated jets in AGNs and GRBs are rather important since they indicate that fast magnetic reconnection can be driven by the kink-instability turbulence and both, govern the transformation of magnetic dominated flows into kinectically dominated ones and provide an efficient way to power and accelerate particles along these relativistic jets by magnetic reconnection and explain the VHE, as predicted in earlier studies. Other concomittant similar MHD  numerical studies to ours above have confirmed  our findings \cite{bromberg16, striani16}.

Finally, it should be mentioned that preliminary tests of $in-situ$ particle acceleration with thousands of test particles in MHD relativistic jets (but considering small sigma values only), have revealed that magnetic reconnection acceleration can be as efficient as shock acceleration in these systems \cite{degouveia15}.

\section{Summary and Conclusions}

In this lecture, we have briefly discussed the potential role of magnetic reconnection and   associated  particle acceleration in the surrounds of  BHs and  relativistic jets, particularly  stressing the importance of turbulence to induce fast reconnection.  

Our main conclusions can be summarized as follows: 
\begin{itemize}

\item First-order Fermi particle acceleration driven by fast magnetic reconnection (numerically tested \cite{gl05, kowal11, kowal12, delvalle16})  can plausibly  explain the observed very high energy  emission  of low-luminous AGNs (LLAGNs)  and microquasars (or GBHs)  as coming from the coronal region around the BH of these sources. The  magnetic  power released in fast reconnection events, calculated as a function of the BH masses (considering a fiducial parametric space depending on the size of the reconnection region and the gas accretion rate),  matches very well with the observed correlation of the radio and gamma-ray luminosities of these sources with their BH masses, spanning over 10 orders of magnitude in mass and power, and tested for more than 230 sources \cite{kadowaki15, singh15}.   

\item Besides, the reconnection acceleration model operating in the core of these sources can also potentially reproduce the observed spectral energy distribution (SED), specially the gamma-ray  band of these sources. This has been tested particularly  for the LLAGNS and GBHs for which emission (or upper limits)  at  the TeV band has been detected (namely, Cen A, Per A, M87 and IC310 among the LLAGNs \cite{khiali16a}, and Cyg X1 and Cyg X3 among the GBHs \cite{khiali15}).  

\item The extragalactic very high energy  neutrino flux recently observed by the IceCube has been also analysed in terms of the magnetic reconnection acceleration model above. In more speculative basis,  it has been argued  that this emission could  be potentially explained as due to core emission from  LLAGNs distributed between 0 and 5.2 redshifts \cite{khiali16b}.
 
\item Finally, we have also discussed fast magnetic reconnection in the framework of magnetically dominated  relativistic jets. 3D  numerical relativistic MHD simulations of rotating tower jets with helical magnetic field subject to current driven kink instability rapidly develop fast turbulent reconnection regions which provide efficient magnetic energy conversion into kinetic energy, and are potential sites for efficient particle acceleretaion and strongly variable non-thermal emission, as required by current observations of GRB and AGN blazar jets \cite{singh16}. 

\end{itemize}

A final note is in order. The  origin of the VHE emission around  BH sources discussed above is still debatable, specially due to the poor resolution of the current gamma-ray detectors.  A core dominated origin, as suggested in this work for the observed  gamma-ray emission of LLAGNs and GBHs  arises as a strong possibility as  long as magnetic activity is significant in the surrounds of the BH, at  the jet launching basis, and gamma-ray aborption via electron-positron pair creation is  negligible, as it seems to be at the viewing angle of these sources. However, in order to ellucidate  this debate, besides extensive global numerical MHD simulations of accretion disk/corona systems around BHs,  we will  need also substantial improvement in the observations, specially in the gamma-ray range. We hope  that with the much larger resolution and sensitivity of the forthcoming gamma-ray CTA observatory  \cite{actis11, acharya13, sol13},   and  with longer times of exposure of  nearby GBHs and LLAGNs, we may collect higher resolution data and more significant information on variability that may help in the determination of the true location of the emission region.

\acknowledgments
We acknowledge support from  the Brazilian agencies FAPESP 
 (2013/10559-5 grant)  and CNPq (306598/2009-4 grant). 
The simulations presented in this lecture have made use of the computing facilities of the Laboratory of Astroinformatics (IAG/USP, NAT/Unicsul),
 whose purchase was made possible by FAPESP (grant 2009/54006-4).

\end{document}